\documentclass[lettersize,journal]{IEEEtran}
\usepackage{amsmath,amsfonts}
\usepackage{algorithmic}
\usepackage{algorithm}
\usepackage{array}
\usepackage[caption=false,font=normalsize,labelfont=sf,textfont=sf]{subfig}
\usepackage{textcomp}
\usepackage{stfloats}
\usepackage{subfig}
\usepackage{url}
\usepackage{verbatim}
\usepackage{graphicx}
\usepackage{cite}
\usepackage{authblk}
\usepackage{amsmath}
\usepackage{multirow}
\usepackage{cleveref} 
\crefname{figure}{fig.}{figs}           

\begin{document}

\title{100 Gbps Quantum-safe IPsec VPN Tunnels over 46~km Deployed Fiber}

\author{Obada~Alia, Albert~Huang, Huan~Luo, Omar~Amer, Marco~Pistoia, and Charles~Lim
\thanks{ The authors are with Global Technology Applied Research at JPMorgan Chase \& Co., e-mail: (obada.alia@jpmchase.com). \\ Manuscript received XXX xx, 2024; revised XXX xx, 2024.}
}

\markboth{XXX XXX XXX, ~Vol.~XX, No.~X, XXX~2024}%
{Shell \MakeLowercase{\textit{et al.}}: A Sample Article Using IEEEtran.cls for IEEE Journals}

\maketitle

\begin{abstract}
We demonstrated for the first time quantum-safe high-speed 100~Gbps site-to-site IPsec tunnels secured using Quantum Key Distribution (QKD) technology. The demonstration was conducted between two JPMorgan Chase Data Centers (DCs) in an air-gapped environment over 46~km of deployed telecom fiber across Singapore achieving 45 days of continuous operation. Two different Virtual Private Network (VPN) tunnel configurations were tested: (1) a QKD-secured VPN tunnel configuration with a maximum throughput of 80~Gbps and (2) a multi-VPN tunnel configuration exhibiting 12 QKD-secured VPN tunnels with a throughput of 8.39~Gbps per tunnel resulting in an aggregated throughput of 99.62~Gbps for all tunnels. For the QKD system performance, we achieved an average Secret Key Rate (SKR) of 7.4~kbps (about 29~AES\nobreakdash-256 keys per second), an average Quantum Bit Error Rate (QBER) of 0.8\% and an average visibility of 98.6\%. We utilized the ETSI\nobreakdash-QKD\nobreakdash-014 REST\nobreakdash-based Application Programming Interface (API) to exchange the QKD generated keys between the key management server in the QKD system and the next-generation firewalls in order to encrypt and decrypt the data. The data was encrypted by the quantum-safe keys using the AES\nobreakdash-256\nobreakdash-GCM cipher suite with a key refresh rate of 120 seconds without affecting the VPN tunnel connectivity and performance. 
\end{abstract}


\section{Introduction}
\label{sec:intro}

Cybersecurity is paramount in our increasingly digital world, especially when dealing with critical infrastructure. Failure to adequately prepare for emerging cyber threats can lead to significant economic losses, breaches of personal rights, loss of critical information, and reputational damages. Recent estimates suggest that by 2025, the cost of cyber-attacks on the global economy will surpass \$10.5~trillion \cite{aiyer2022new}. This staggering figure underscores the need to position cybersecurity as a strategic priority at individual, organizational, and governmental levels. 

The development of large-scale quantum computers currently poses one of the most significant and large-scale threats to cybersecurity standards, compelling users and practitioners to rethink the way that they protect sensitive infrastructure and information. As is widely known by now, large-scale quantum computers, if realized, could lead to the breakage of today's most prevalent cryptographic standards such as RSA and Diffie-Hellman. Today these cryptographic standards are enablers for authenticity, confidentiality, and integrity, and thus constitute the bedrock for securing digital products and services. Indeed, even prior to the arrival of large-scale quantum computers, bad actors could already be pilfering data and storing it until decryption becomes feasible; an attack called harvest-now decrypt-later (HNDL). It is thus of critical importance to prioritize addressing the quantum threat in order to limit exposure to eventual quantum attackers.

In response to the threat, the National Institute of Standards and Technology (NIST) advocates an accelerated transition to Post-Quantum Cryptography (PQC) to protect critical infrastructure against quantum computers. Despite uncertainties about the ultimate target, organizations must initiate the lengthy shift towards PQC now, employing a phased approach to facilitate migration at each stage of the journey. Another promising solution for a quantum-safe infrastructure is to deploy \emph{Quantum Key Distribution} (QKD), which exploits the laws of quantum mechanics to distribute secret symmetric keys between authenticated users embedded in an untrusted optical network~\cite{scarani2009security}. These secret keys can be used for data encryption, which protects sensitive information against unauthorized parties.

The scale and viability of QKD networks has grown rapidly in recent years, evolving from experimental lab-based test-beds dating back to 2004, to field trials over deployed optical fibers around the world \cite{elliott2005current, dynes2019cambridge}. The Cambridge QKD network is an example of a city-wide metropolitan QKD network operating on deployed fibers already populated with high-bandwidth data traffic enabled using Dense Wavelength Division Multiplexing (DWDM) technology \cite{dynes2019cambridge}. The network reported a stable and consistent operation for over two years across its three links with a constant secret key rate throughout the whole operation. Other networks such as the Madrid QKD network implemented a software-defined networking (SDN) architecture with an SDN controller capable of optimizing and integrating quantum and classical technologies in a production environment \cite{aguado2019engineering}. Furthermore, long-haul QKD networks have been implemented across the globe, reaching a distance of 2,000~km for trusted-node fiber-based networks and 4,600~km using satellite technology \cite{liao2018satellite}. In 2022, BT Group and Toshiba built the world's first commercial quantum-secured metro network across London, which is currently being used by customers from different industries including financial institutions. 

Simultaneously, QKD technology has made enormous progress in terms of its practical usability, in part due to its integration and compatibility with different layers of the Open Systems Interconnection (OSI) model including the physical layer (layer 1), data link layer (layer 2) and the network layer (layer 3). More concretely, it has been integrated successfully within protocols at the aforementioned communication layers, including optical encryption (OTNsec)~\cite{pistoia2023paving}, Media Access Control security (MACsec) \cite{cho2021using}, and Internet Protocol security (IPsec). 

IPsec plays a crucial role in providing robust security features for the Internet Protocol (IP) at layer 3, including confidentiality, data integrity, access control, and data source authentication, to IP datagrams. The Internet Key Exchange (IKE) protocol IKEv2 serves as the core protocol of IPsec, with its primary objective being to establish and maintain shared security parameters and authenticated keys between two IPsec endpoints \cite{rfc7296}. IPsec implementations are of particular interest due to their commonplace use in enabling Virtual Private Network (VPN) services. VPNs create a secure virtual network over existing public networks, such as the Internet, and thus are central to the security of modern enterprise networks and other digital platforms. Indeed, VPNs are critical in enabling remote workers to access the private networks of their organizations, and it is even common for VPNs to be deployed on dedicated private fibers as part of multi-tenant solutions. Given the ubiquity and importance of IPsec in robust cybersecurity strategies, it is of significant interest to integrate QKD technology further into the high-speed IPsec networks and their underlying infrastructures \cite{Sami2020}.

Integrating QKD with IPsec significantly enhances network security by combining robust encryption with quantum-resistant key exchange, ensuring the confidentiality and integrity of sensitive data with future-proofing against emerging classical and quantum attackers in an evolving threat landscape \cite{zhang2022future}. Towards that end, the European Telecommunications Standards Institute (ETSI) has already published  ETSI\nobreakdash-QKD\nobreakdash-014, a standard describing a REST-based key delivery Application Programming Interface (API) for QKD systems to standardize the integration with classical technologies across multiple layers. The ETSI key delivery API has already been adopted by QKD vendors such as Toshiba and ID~Quantique and by classical encryptor vendors such as Ciena, Juniper, Fortinet, and Thales.

In this work, we present an operational quantum-safe network achieving 100~Gbps QKD-secured IPsec VPN tunnels over 46~km of deployed telecom fiber with over 45 days of continuous operation. This paper is organised as follows: \Cref{sec:setup} describes the design architecture of the network and highlights the equipment and software configuration used in this demonstration. \Cref{sec:results} discusses the experimental results of the QKD system, traffic generator and VPN tunnels throughput. Finally, the conclusion is presented in \Cref{sec:conclusion}.

\section{Field trial Setup}
\label{sec:setup}

\begin{figure*}[t!]
\centering
\includegraphics[width=\linewidth]{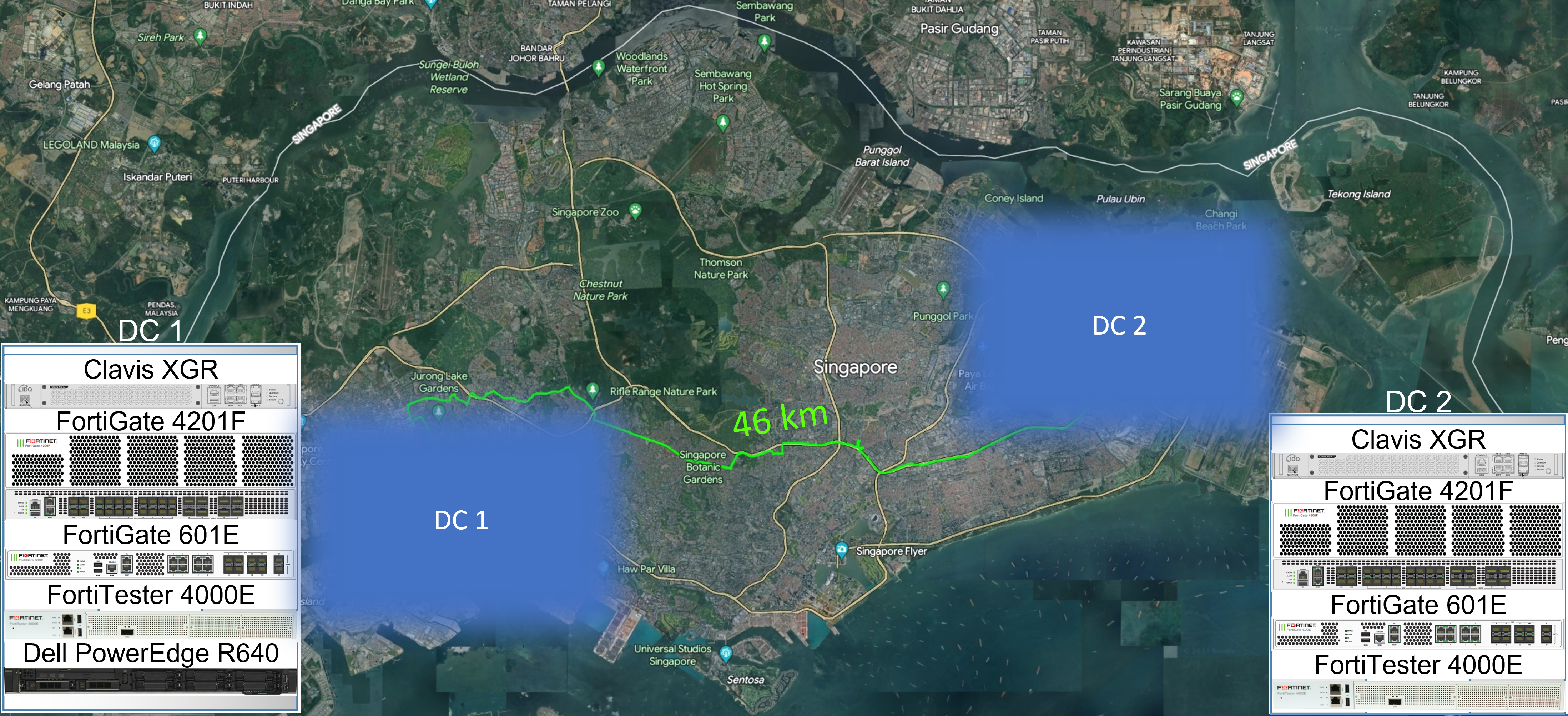}
\caption{A point-to-point 46~km QKD link in a metropolitan network across Singapore. Two nodes denoted DC 1 and DC 2 are connected directly using underground telecom single-mode optical fibers (green line). The rack figures depict the equipment deployed in each data center. }
\label{fig:map}
\end{figure*}

\subsection{Quantum Key Distribution System}

In this trial, we used the ID~Quantique Clavis XGR QKD system. This QKD system, occupying a single rack unit (1U), operates within the C-band (specifically, channel 32 in the 100 GHz ITU-T grid) and has a loss budget of 24~dB, facilitating transmission over 120~km of single-mode fiber. The Clavis XGR QKD system implements the decoy state four-state BB84 protocol \cite{Hwang2003,Lo2005,Lim2014} to encode classical bits into quantum bits (qubits) using time-bin encoding. The decoy state four-state BB84 protocol employs two bases: time-bins (Z-basis) and phase (X-basis). The transmitter, XGR-Alice (XGR-A), features a pulsed laser operating at 1551.72~nm (C-band). This laser beam undergoes subsequent modulation in both intensity and phase to generate optical pulses, each representing a value of zero and one. These pulses are then attenuated to the single-photon level. Traveling from the transmitter (XGR-A) through the quantum channel, they reach the receiver, XGR-Bob (XGR-B), where detection occurs. Bits 0 and 1 are observed either on the Z-basis detector for intensity modulation or on the X-basis detector for phase modulation. Notably, the Secret Key Rate (SKR) calculation is solely based on measurements made by the Z-basis detector. The Clavis XGR QKD system incorporates the ID~Quantique Clarion KX Quantum Key Management System (KMS) which oversees key requests and transfers between QKD systems and external encryptors. The distribution of keys to encryptors or any key consumer is executed through the secured ETSI\nobreakdash-QKD\nobreakdash-014 REST API or through proprietary interfaces developed in collaboration with prominent vendors, such as the Secure Key Integration Protocol (SKIP) used by Cisco encryptors. 

\subsection{Hardware equipment and fiber connections}
The equipment used in the field trial are shown in \Cref{fig:map}. In addition to the Clavis XGR QKD system, we used the following equipment: a central compute server (Dell PowerEdge R640) to access and configure the equipment, a management switch (FortiGate 601E) to combine the management traffic of all equipment, a next-generation firewall (FortiGate 4201F) to encrypt and decrypt the data, and a traffic generator (FortiTester 4000E) to generate data traffic for performance testing and validating network security infrastructure. All of the used equipment are available commercially.

We also used five fiber pairs consisting of 2xdark fiber pairs, 2x1G telecom circuits, and 1x100G telecom circuits as shown in \Cref{fig:testbed}. The fiber pairs connect the two data centers (DCs), passing through three telecom exchange buildings along the path from DC 1 to DC 2 (green line in \Cref{fig:map}). The fiber spans 46~km with an optical insertion loss of 11.15~dB and an average loss of 0.242~dB/km. As shown in \Cref{fig:testbed}, the two dark fibers are used to connect the quantum channel (black) and service channel (purple) of the QKD system. The two 1G telecom circuits are used to connect the KMS channel (grey) and management channel (green), and the one 100G telecom circuit is used for the encrypted data channel (blue). We note that the 1G and 100G telecom circuits could have been replaced with dark fiber pairs if desired. Moreover, while out of the scope this demonstration, in principle all the channels could be mutliplexed using DWDM technology into a single dark fiber pair, as was discussed briefly in \Cref{sec:intro}.

\subsection{Experimental Testbed}

\begin{figure*}[t!]
\centering
\includegraphics[width=\linewidth]{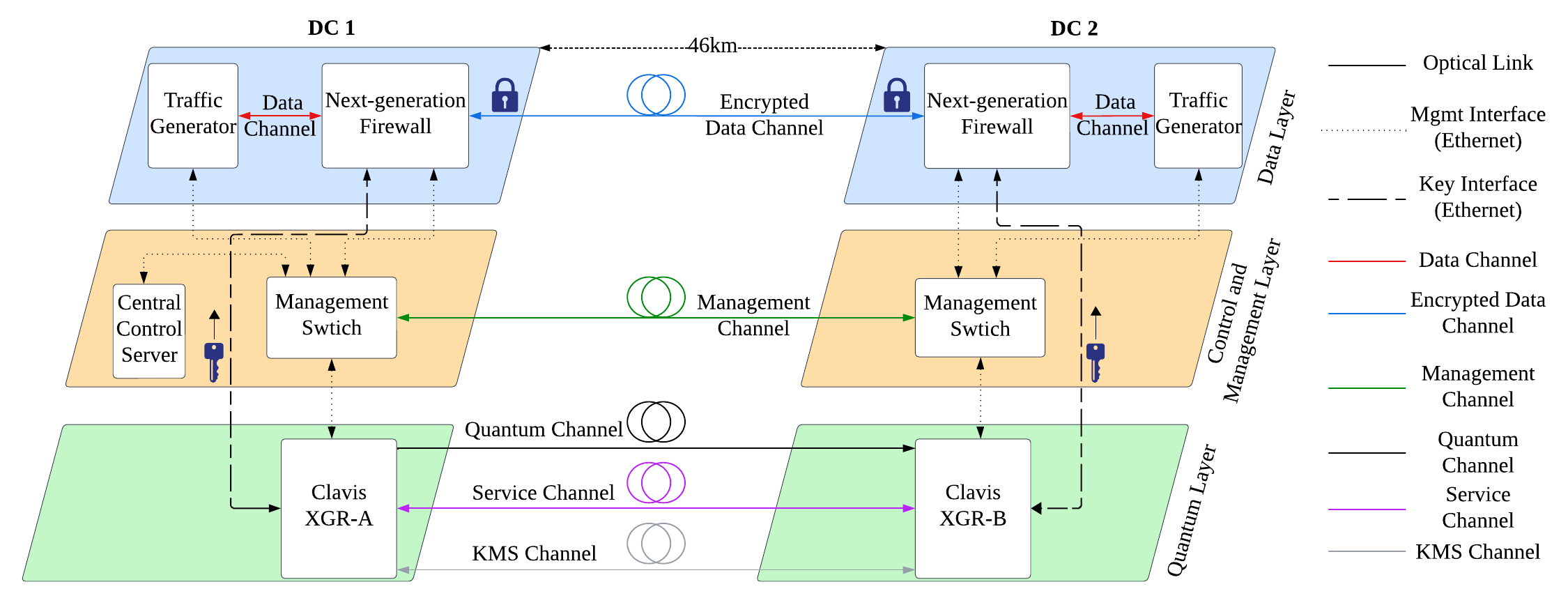}
\caption{Experimental Testbed. The testbed includes three different layers, the quantum later (green), the control and management layer (orange), and the data layer (blue).}
\label{fig:testbed}
\end{figure*}

The experimental setup used in the field trial is shown in \Cref{fig:testbed}. As mentioned before, the QKD systems were connected using two dark fiber pairs and one 1G telecom circuit. One of the dark fibers was used for the uni-directional quantum channel which connected XGR-A directly to XGR-B. The other dark fiber was used for the service channel which implemented the proprietary IDQ4P communication protocol to enable the clock synchronization and post-processing of the QKD systems. DWDM 2.7G SFP transceivers were used for the service channel to handle the traffic and to enable classical bi-directional communication between the QKD systems. The KMS channel arbitrates the key distribution between the QKD system and key consumers, and was connected using a 1G SFP transceiver via the 1G telecom circuit. Similarly, a 1G SFP transceiver and a 1G telecom circuit were used for the management channel to connect the two management switches. The management switches in both DC 1 and DC 2 were utilized to manage and monitor the equipment and to combine the management traffic of the QKD systems, NGFWs, compute server, and traffic generator in both DCs via the management interface links (dotted lines in \Cref{fig:testbed}). Finally, 100G Quad SFP28 (QSFP28) transceivers was used to transmit the unencrypted data channel (red) between the traffic generator and the NGFW. 100G QSFP28 transceivers were also used to transmit the encrypted data between the NGFWs in both DCs across the data channel (blue) via a high-speed 100G circuit. \\

In this demonstration, we used a modified version of the RFC 7296 IKEv2 implementation \cite{rfc7296}, in which the Child Security Association (Child~SA) establishment process allows for the use of QKD-keys as an input to the key derivation process. We describe below the relevant portions of the both the larger IKE Security Association (IKE~SA) establishment process and the Child~SA. Among other functions, the IKE~SA is responsible for mutually authenticating IPsec endpoints, establishing shared secret information to be used in the establishment of subsequent Child~SAs, and negotiating cryptographic algorithms for integrity protection, key exchange, encryption and decryption, and key derivation. The Child~SAs, meanwhile, are used by endpoints with an established IKE~SA to achieve the desired integrity and confidentiality functionalities.


As part of the IKE~SA establishment process, a seed for later use is established using the pseudorandom function as below:
\newcommand\numberthis{\addtocounter{equation}{1}\tag{\theequation}}

\begin{align*}
\mathsf{KEY_{seed} = prf(Ni | Nr, g^{ir})} 
\numberthis \label{eq:Key_seed}
\end{align*}


where, $\mathsf{g^{ir}}$ is a shared secret established during an ephemeral Diffie-Hellman exchange, and $\mathsf{Ni}$ and $\mathsf{Nr}$ are nonces shared earlier in the IKE~SA establishment process. 

The shared seed is then used as an input to a length-extended variant of the same pseudorandom function, denoted $\mathsf{prf_+}$, in order to derive seven separate keys for use in a variety of functions as part of the greater IKE process. Most relevant to us is $\mathsf{SK_d}$, which is used to derive new keys for Child~SAs established under the IKE~SA. 


\begin{align*}
    \mathsf{SK_d | SK_{ai} | SK_{ar} | SK_{ei} | SK_{er} | SK_{pi} | SK_{pr}} \\
    = \mathsf {prf_+ (KEY_{seed}, Ni | Nr | SPIi | SPIr)}
\numberthis \label{SA_Eq}
\end{align*}


Along with $\mathsf{KEY_{seed}}$, a number of other inputs are used to establish the keys, as shown above in \Cref{SA_Eq}. In addition to the previously used nonces, $\mathsf{SPIi}$ and $\mathsf{SPIr}$ are unique values established earlier in the IKE~SA flow. 


For the Child~SAs, we use a modified key derivation process, shown below, that that uses the QKD keys as a fresh source of entropy rather than a Diffie-Hellman key exchange:

\begin{align*}
\mathsf{KEY_{material} = prf_{+}(SK_d, QKD_{Key} | Ni | Nr)}
\numberthis \label{Child_SA_Eq}
\end{align*}
where $\mathsf{prf_+}$ is again a variant of the negotiated pseudorandom function with length-extension achieved, here and where previously used, through iteration as dictated in \cite{rfc7296}.


Ultimately, the keys established by subsequent Child~SAs are derived from entropy sources tied to both the initial Diffie-Hellman exchange conducted during the exchange (specifically, $SK_d)$, and from the independently generated QKD key. As such, the resulting keys cannot be learned even by an attacker who may capture and break the first Diffie-Hellman based exchange. 

To enable this enhancement, QKD keys are transmitted via the Key interface link (dashed line in \Cref{fig:testbed}) from the QKD system to the NGFW using the ETSI\nobreakdash-QKD\nobreakdash-014 API. The QKD keys are utilized in the manner described above to establish encryption and decryption keys for the end points. Finally, the NGFW encrypts the data stream generated by the traffic generator in the data channel using IPsec with AES\nobreakdash-256\nobreakdash-GCM cipher suite using the derived quantum-safe keys. 

\section{Results}
\label{sec:results}

In this section, we show the QKD system stability by analysing the SKR, QBER and visibility. We also share parameters across different layer for the un-encrypted data generated by the FortiTester. Finally, we provide the VPN tunnels throughput (encrypted data channel) for both the one and multi tunnel configurations.

\begin{figure}[t!]
\centering
\includegraphics[width=\linewidth]{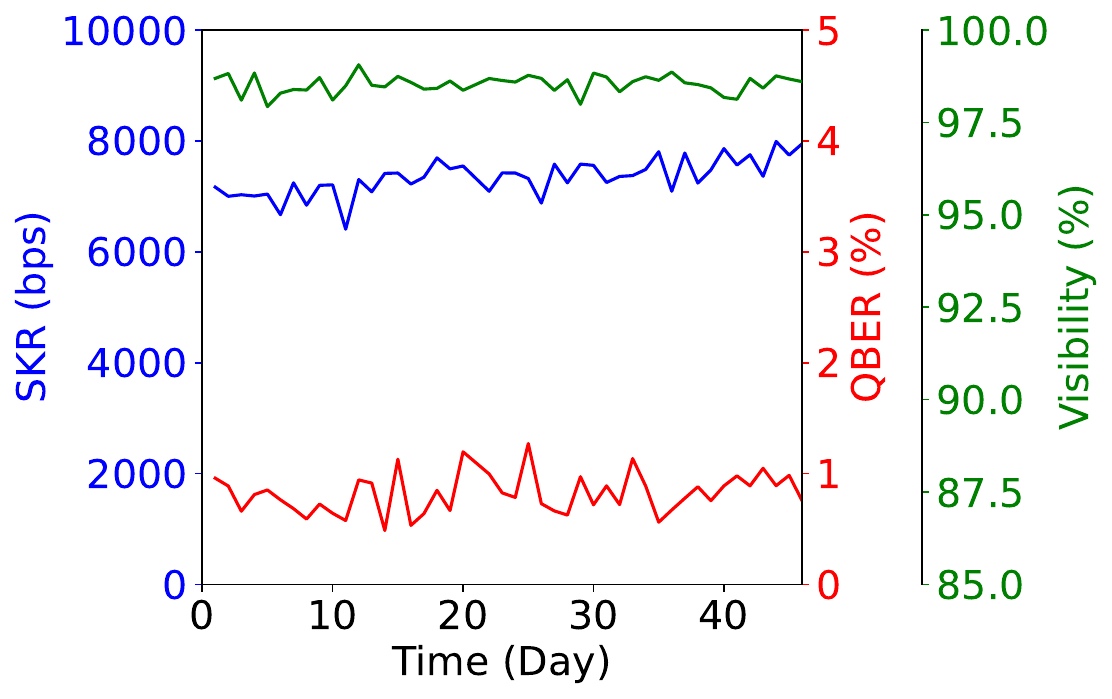}
\caption{QKD system performance and stability over 45 days; Secret key rate (solid blue), Quantum bit error rate (dotted red), and Visibility (dashed green).}
\label{fig:QKD_performance}
\end{figure}


\subsection{QKD System Performance and Stability}
The performance and stability of the QKD system were evaluated by measuring the SKR, QBER, and visibility over 45 days between the two data centers. As shown in \Cref{fig:QKD_performance} and based on over 290,000 data points, we observed an average SKR of 7.4~kbps (about 29~AES\nobreakdash-256 keys per second), an average QBER of 0.8\% and an average visibility of 98.6\%. 

\subsection{Traffic Generator Data}
\Cref{tab:tester} show the traffic generator measurements for the un-encrypted data channel (red in \Cref{fig:testbed}) in one second as measured by the FortiTester traffic generator. We collected data for traffic at layers 2, 3 and 4. For layer 2, we present the bandwidth in Mbps sent (Tx) and received (Rx), as well as the total bandwidth. For layer 3 we record instead the total number of packets sent and received, as well as any packets dropped in either direction. Finally, for layer 4, we show the number of concurrent UDP communication streams that can be supported simultaneously. 


We ran a stress test for 24 hours, however, we only included the measurement for one second in \Cref{tab:tester}. As shown in \Cref{tab:tester} starting from the bottom up, we recorded a bandwidth of 100~Gbps for both the client and the server at layer 2. In layer 3, over 1.5 Million packets were sent and received in one second by both the client and receiver with zero packets being dropped. Finally, in layer 4 over 12,000 UDP communication streams were handled simultaneously every second by the NGFWs. Over the full 24 hours, 15~Pb of data were sent at layer 2, and at layer 3, a total of 131 Billion packets were sent and fewer than 20,000 packets were dropped for a drop rate of less than 2$\times$10$^{-7}$.

\begin{table}[h]
\centering
\caption{Traffic generator (FortiTester 4000E) data}
\label{tab:tester}
\begin{tabular}{cccc}
                                                                             &                    & Client  & Server \\ & & (per second)  & (per second)\\
\hline
Layer 4                                                                      & UDPv4 Concurrency & 12,288     & 12,288     \\
\hline
\multirow{6}{*}{\begin{tabular}[c]{@{}c@{}}Layer 3\\ (Packets)\end{tabular}} & Tx                 & 1,525,632  & 1,525,632  \\
                                                                             & Rx                 & 1,515,660  & 1,515,712  \\
                                                                             & Dropped Tx        & 0          & 0          \\
                                                                             & Dropped Rx        & 0          & 0          \\
\hline
\multirow{3}{*}{\begin{tabular}[c]{@{}c@{}}Layer 2\\ (Mbps)\end{tabular}}    & Tx                 & 100,227    & 100,227    \\
                                                                             & Rx                 & 99,572     & 99,576     \\
                                                                             & Bandwidth          & 199,800    & 199,804    \\
\hline
\end{tabular}
\end{table}

\subsection{IPsec VPN Tunnels Throughput}

\begin{figure*}[t]
\centering
\begin{tabular}{cc}
\subfloat[\label{fig:one_VPN_results}] { \includegraphics[width=.45\textwidth]{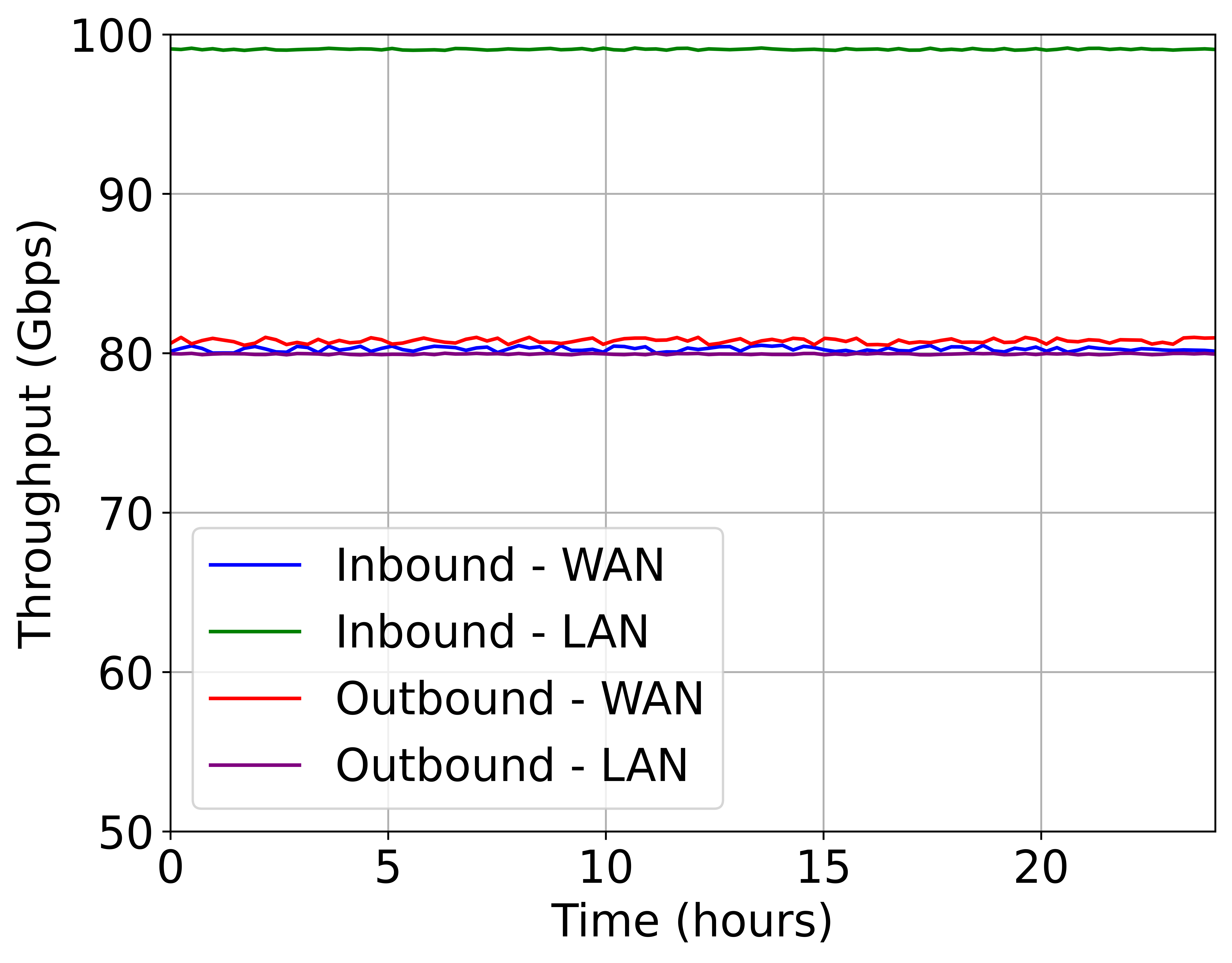} } & 
\subfloat[\label{fig:Multi_vpn_results}] { \includegraphics[width=.45\textwidth]{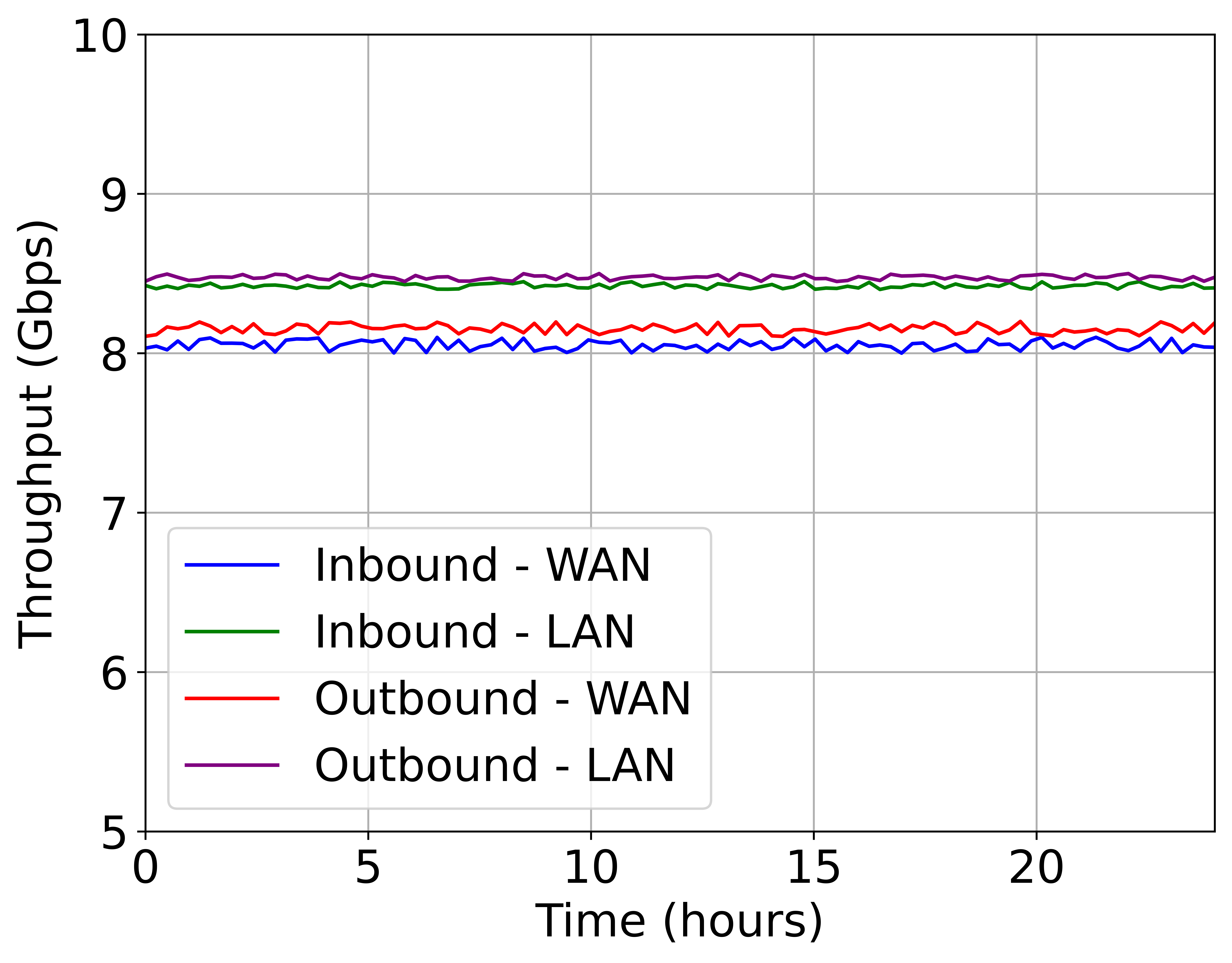}
 }
\end{tabular}
\caption{IPsec VPN tunnels inbound and outbound throughput for two configurations during a 24-hour stress test. (a) The throughput of single VPN tunnel configuration. (b) The throughput of two VPN tunnels in a multi VPN tunnel configuration.}
\label{fig:NGFW_data}
\end{figure*}

We measured the throughput of inbound and outbound traffic during the 24-hour stress test for two different VPN configurations. \Cref{fig:NGFW_data}a shows the throughput of a single tunnel in both the wide-area network (WAN) link (red and blue), carrying encrypted data via the VPN tunnel between the NGFWs, and the local area network (LAN) link (purple and green) carrying unencrypted data between the traffic generator and the NGFW. We observed outbound throughput for the WAN (red) and LAN (purple) links of 80.45~Gbps and 79.92~Gbps respectively, and inbound throughput for the WAN (blue) and LAN (green) links of 80.45~Gbps and 99.14~Gbps respectively. We also achieved a peak VPN throughput of over 90~Gbps for a single VPN tunnel configuration by using the iPerf software to generate UDP data streams and measure the throughput of the network. However, the test was conducted over a short period of time over a few minutes.

\Cref{fig:NGFW_data}b shows results for the multi-VPN tunnel configuration, where each VPN tunnel had its own encryption key that was generated from separate QKD keys and is therefore independent from the encryption keys used by the other VPN tunnels. \Cref{fig:NGFW_data}b shows 2 out of the 12 VPN tunnels that were tested simultaneously in the multi VPN tunnel configuration, with an average inbound and outbound throughput of 8.33~Gbps and 8.31~Gbps respectively. The aggregated inbound and outbound VPN tunnel throughput of all 12 VPN tunnels was 99.62~Gbps in both cases. To put that into perspective, public cloud providers usually provide site-to-site VPN connection with just two tunnels, with each tunnel supporting a maximum throughput of up to 1.25~Gbps. Compared to the public cloud provider VPN throughput, we achieved 64 times the throughput in the single VPN tunnel configuration and 6 times the throughput for each VPN tunnel in the 12 VPN tunnel configuration. It is clear from \Cref{fig:NGFW_data} that the traffic was consistent throughout the full 24-hour testing period for both configurations with a stable operation of the FortiGate 4201F NGFW.


The encryption key refresh rate for the VPN tunnel was 120 seconds, which is much lower than the National Institute of Standards and Technology (NIST) recommended cryptoperiod of a day or a week for refreshing symmetric data-encryption keys \cite{elaine2020recommendation}. Every 120 seconds, a new QKD\nobreakdash-generated key is sent from the QKD system to the NGFW before the old key expires. As mentioned before, we observed an SKR of 7.4~kbps resulting in about 29~AES\nobreakdash-256 keys per second. In other words, the system generates 3360 AES\nobreakdash-256 keys between refreshes, with only~0.03\% of the keys being consumed during each key refresh in the single-VPN tunnel configuration, and ~0.3\% being consumed in the multi-VPN tunnel configuration. Indeed, even if we were to refresh the key every second to encrypt 100~Gbit of data, we would have only consumed~3.6\% of the available keys. We note that in our demonstration, a single key was used for the encryption of 1.5~TByte of data (100~Gbps x 120~seconds), which is within the recommendation of the AES\nobreakdash-256\nobreakdash-GCM and the IKEv2 protocol.

\section{Conclusion}
\label{sec:conclusion}
We demonstrated quantum-safe 100~Gbps IPsec VPN tunnels over 46~km between two data centers with over 45~days of continuous operation. The QKD system was stable throughout the testing period and we achieved an average SKR of 7.4~kbps (about 29~AES\nobreakdash-256 keys per second), an average QBER of 0.8\% and an average visibility of 98.6\%. We tested two different configurations with QKD integration, one with single VPN tunnel, and one with twelve concurrent VPN tunnels. We managed to transmit 131~Billion packets with bandwidth of about 15~Pbits over a period of 24~hours (around~173~Gbps) via the VPN tunnel using QKD derived secret keys to encrypt the data with the AES\nobreakdash-256\nobreakdash-GCM cipher suite. We achieved a total throughput of 80~Gbps for the one tunnel configuration and a total throughput 99.62~Gbps for the multi tunnel configuration. Such throughput supports site-to-site VPN tunnels for different real-world industry use cases, such as between on-premises locations (DC to DC, or DC to office), between on-premises locations and cloud service providers, and to secure traffic between radio access networks (RANs) and backbone core networks. We utilized ETSI\nobreakdash-QKD\nobreakdash-014 API to provide QKD keys to the NGFW for the derivation of encryption keys with an encryption key refresh rate of 120~seconds and a QKD key consumption rate of~0.03\% and 0.3\% for the single and multi configurations respectively. The stability of the VPN tunnels, overabundance of QKD keys, and unimpeded throughput of the VPN tunnels supports the finding that in such deployments, QKD integration achieves enhanced security without sacrificing performance. 

%

\section*{Disclaimer}
{This paper was prepared for informational purposes by the Global Technology Applied Research Center of JPMorgan Chase \& Co. This paper is not a product of the Research Department of JPMorgan Chase \& Co. or its affiliates. Neither JPMorgan Chase \& Co. nor any of its affiliates makes any explicit or implied representation or warranty and none of them accept any liability in connection with this paper, including, without limitation, with respect to the completeness, accuracy, or reliability of the information contained herein and the potential legal, compliance, tax, or accounting effects thereof. This document is not intended as investment research or investment advice, or as a recommendation, offer
, or solicitation for the purchase or sale of any security, financial instrument, financial product or service, or to be used in any way for evaluating the merits of participating in any transaction.}

\bibliographystyle{IEEEtran}
\bibliography{sample}

\section{Biography Section}
\begin{IEEEbiography}[{\includegraphics[width=1in,height=1.25in,clip,keepaspectratio]{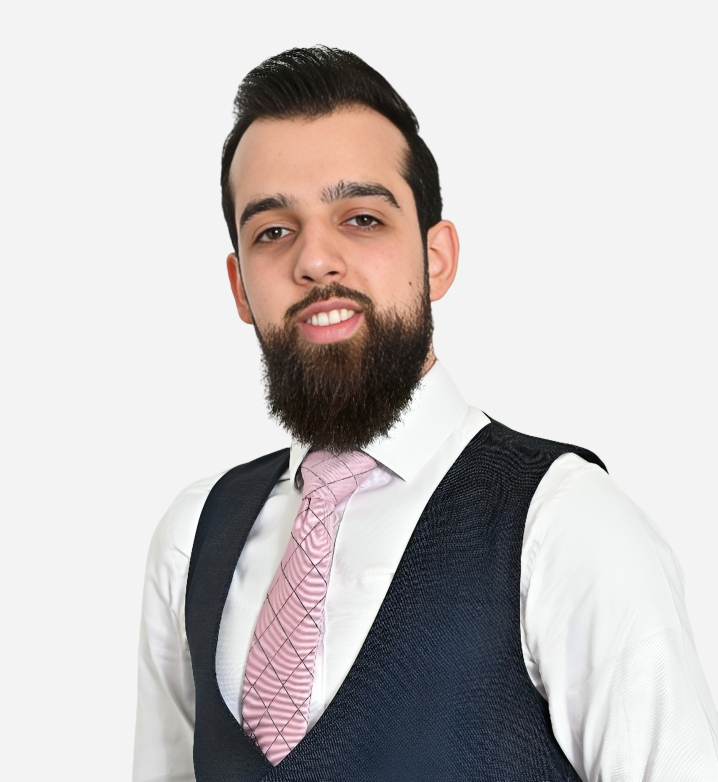}}]{Obada Alia, PhD, is a Senior Applied Research Associate in Quantum Communications and Cryptography at the Global Technology Applied Research Center of JPMorgan Chase \& Co. His role explores the use of quantum cryptography and next-generation communication networks in the financial services industry. Previously, he held the position of a Senior Research Associate at the University of Bristol, where he worked on the coexistence of quantum and classical technologies in dynamic networks. Obada holds a PhD in Quantum Communication from the University of Bristol, along with master's and bachelor's degrees in Telecommunications, Engineering and Electrical and Electronics Engineering from the same university. His doctoral research focused on the integration of quantum and classical technologies within telecommunication infrastructure. Throughout his academic journey, Obada actively contributed to prominent projects, including the UK Quantum Communication Hub 1\&2 and the H2020 UNIQORN project}
\end{IEEEbiography}

\begin{IEEEbiography}[{\includegraphics[width=1in,height=1.25in,clip,keepaspectratio]{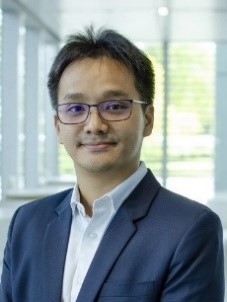}}]{Albert Huang holds the position of Executive Director in the Global Technology Applied Research (GTAR) Quantum Communications and Cryptography team, leading the engineering efforts within the group. He joined JPMorgan Chase \& Co. in September 2022 as part of the pioneering team based in Singapore. Previously, Albert held the position of ecosystem builder in the Quantum Engineering Programme (QEP) office from 2021 to 2022, successfully onboarding 15 partners in formation of the National Quantum Safe Network. Albert started his career at the Centre for Strategic Infocomm Technologies (CSIT) before joining SwissQual (Asia Pacific) which was acquired by Rohde and Schwarz (R\&S). Within R\&S Regional Headquarters Singapore, Albert held various positions, his last role being Head of Operations for Mobile Network Testing Asia Pacific Support. Albert was also briefly with the Ministry of Communications and Information as Assistant Director Technology Strategy and Planning.}
\end{IEEEbiography}

\begin{IEEEbiography}[{\includegraphics[width=1in,height=1.25in,clip,keepaspectratio]{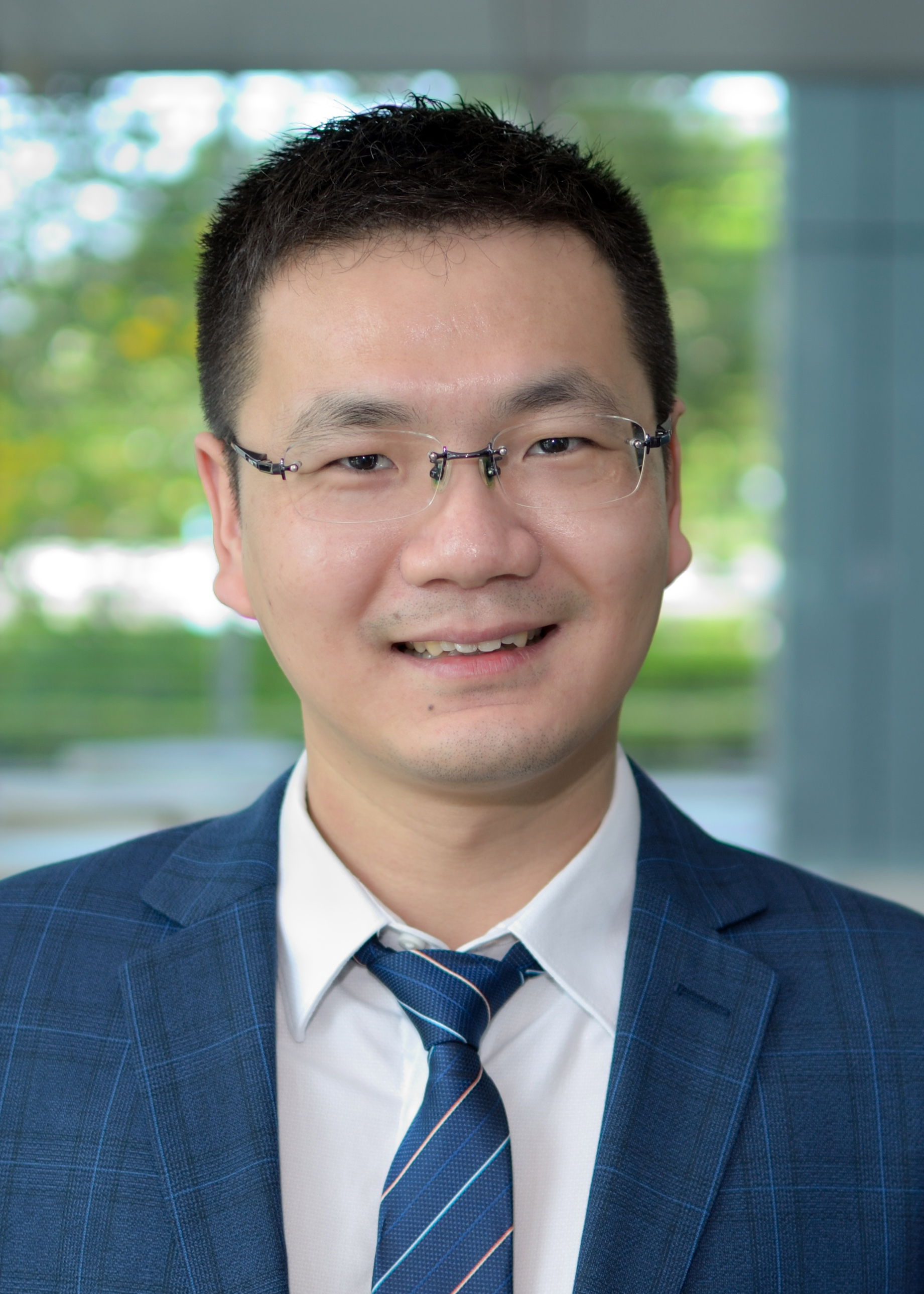}}]{Huan Luo has been infrastructure engineer in Quantum Communications and Cryptography at the Global Technology Applied Research Center of JPMorgan Chase \& Co. since Mar 2023. He has helped complete proof of implementation of Quantum Key Distribution within JPMC Data Center environment. He is experienced in designing network architecture, implementing networking solutions and configuring network protocols. Huan holds a Bachelor of Engineering in Computer Sciences from Nanyang Technological University in Singapore.}
\end{IEEEbiography}

\begin{IEEEbiography}[{\includegraphics[width=1in,height=1.25in,clip,keepaspectratio]{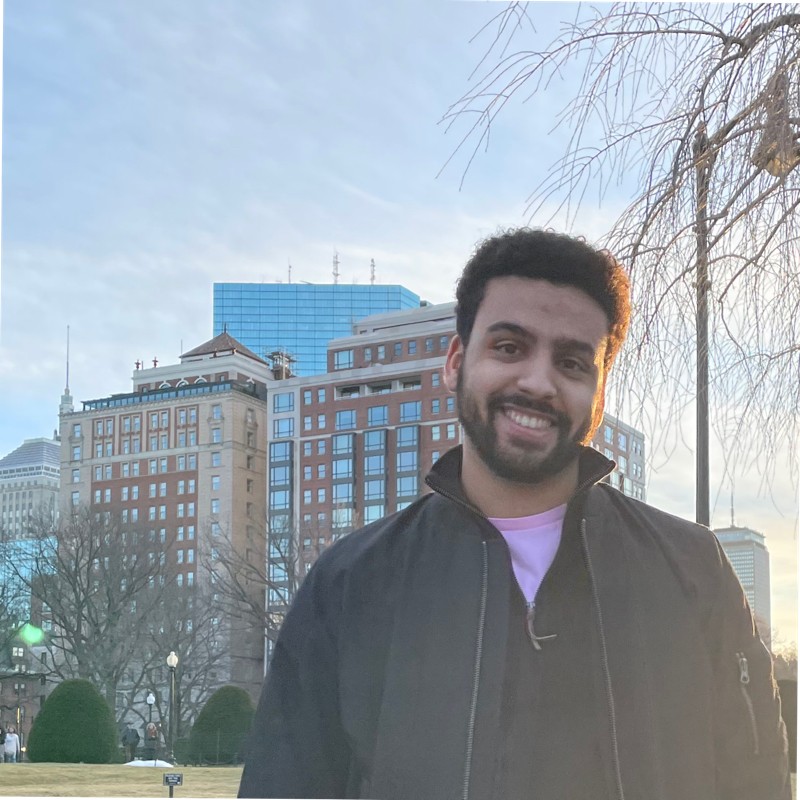}}]{Omar Amer, PhD, is a Vice President of Applied Research at JPMorgan Chase \& Co. working towards making practical the use of QKD in the financial services industry. He received his PhD in Computer Science from the University of Connecticut in 2022, where he studied semi-quantum key distribution protocols, routing in hybrid QKD networks, and practical quantum key distribution.}
\end{IEEEbiography}

\begin{IEEEbiography}[{\includegraphics[width=1in,height=1.25in,clip,keepaspectratio]{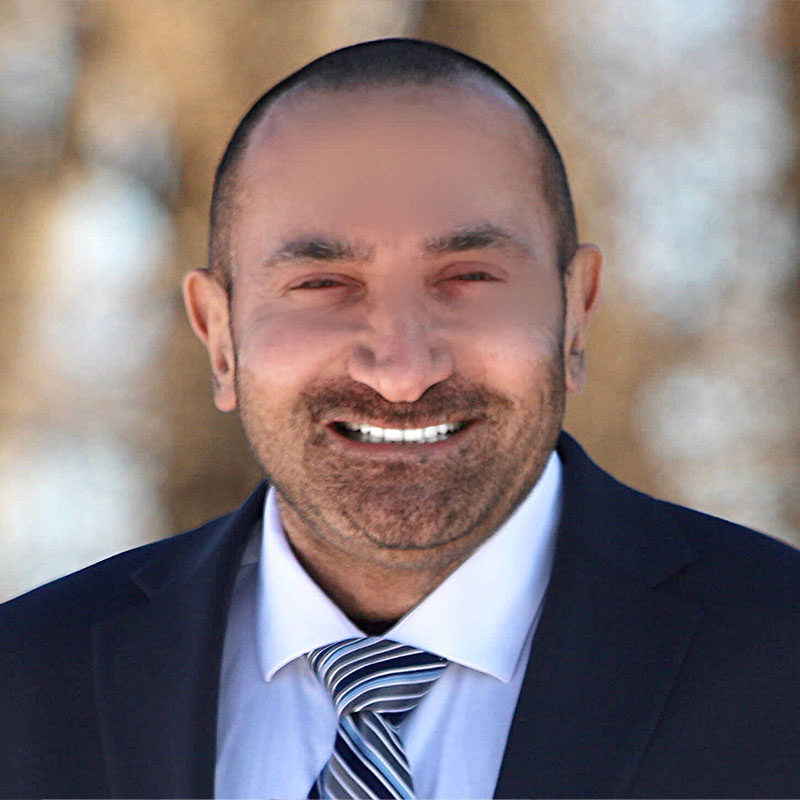}}]{Marco Pistoia, PhD, is Managing Director, Distinguished Engineer and Head of JPMorgan Chase's Global Technology Applied Research (formerly Future Lab for Applied Research and Engineering), where he leads research in Quantum Computing, Quantum Communication, Cloud Networking, Augmented and Virtual Reality (AR/VR), Internet of Things (IoT), and Blockchain and Cryptography. He joined JPMorgan Chase in January 2020.  Formerly, he was a Senior Manager, Distinguished Research Staff Member and Master Inventor at the IBM Thomas J. Watson Research Center in New York, where he managed an international team of researchers responsible for Quantum Computing Algorithms and Applications. He is the inventor of over 250 patents, granted by the U.S. Patent and Trademark Office, and over 300 patent-pending applications. Over 40 of his patents are in the area of Quantum Computing.}
\end{IEEEbiography}

\begin{IEEEbiography}[{\includegraphics[width=1in,height=1.25in,clip,keepaspectratio]{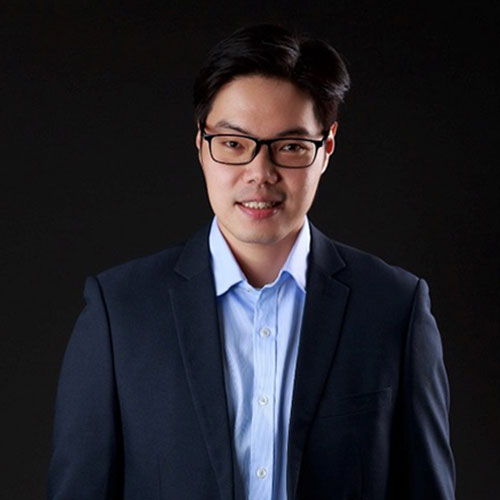}}]{Charles Lim, Ph.D. is Managing Director and the Global Head of Quantum Communications and Cryptography at JPMorgan Chase \& Co. His role explores the use of quantum information technology in cryptography and next-generation communication networks. Dr. Lim is a recipient of the highly prestigious National Research Foundation (Singapore) Fellowship and an Associate Professor (on leave) at the National University of Singapore (NUS). Dr. Lim's research is focused on fundamental quantum technologies that will enable the next-generation communication networks. He is currently leading an international effort within the ISO framework to standardize security requirements for Quantum Key Distribution. In 2021, he was called upon to lead Singapore’s national testbed for quantum-safe digital solutions. He is currently the co-chair of World Economic Forum’s Global Future Council on the Future of Quantum Economy.}
\end{IEEEbiography}

\vfill

\end{document}